# An Advanced Certain Trust Model Using Fuzzy Logic and Probabilistic Logic theory

Kawser Wazed Nafi[1,4], Tonny Shekha Kar[1,4], Md. Amjad Hossain[2,4], M.M.A Hashem[3,4]
[1]Lecturer, Computer Science and Engineering, Stamford University Bangladesh,
[2]Assistant Professor, Computer Science and Engineering,
[3]Professor, Computer Science and Engineering,
[4]Khulna University of Engineering and Technology, Bangladesh

*Abstract*—Trustworthiness especially for service oriented system is very important topic now a day in IT field of the whole world. Certain Trust Model depends on some certain values given by experts and developers. Here, main parameters for calculating trust are certainty and average rating. In this paper we have proposed an Extension of Certain Trust Model, mainly the representation portion based on probabilistic logic and fuzzy logic. This extended model can be applied in a system like cloud computing, internet, website, e-commerce, etc. to ensure trustworthiness of these platforms. The model uses the concept of fuzzy logic to add fuzziness with certainty and average rating to calculate the trustworthiness of a system more accurately. We have proposed two new parameters - trust T and behavioral probability P, which will help both the users and the developers of the system to understand its present condition easily. The linguistic variables are defined for both T and P and then these variables are implemented in our laboratory to verify the proposed trust model. We represent the trustworthiness of test system for two cases of evidence value using Fuzzy Associative Memory (FAM). We use inference rules and defuzzification method for verifying the model.

I. INTRODUCTION

TRUST is a well-known concept in everyday life and often serves as a basis for making decisions in complex situations. There are numerous approaches for modeling trust concept in different research fields of computer science, e.g., virtual organizations, mobile and P2P networks, and E-Commerce. The sociologist Diego Gambetta has provided a definition, which is currently shared or at least adopted by many researchers. According to him "Trust is a particular level of the subjective probability with which an agent assesses that another agent or group of agents will perform a particular action, both before he can monitor such action and in a context in which it affects its own action" [1].

The trustworthiness of the overall system depends on the following things:

*a) The trustworthiness of the subsystems and atomic component independently from how these trust values are assessed.*

*b) Information on how the system combines its subsystems and components.*

*c) The knowledge about which subsystems and components are redundant.*

Therefore, a major challenge of serving trust for the overall system is needed to consider that in real world applications the information about the trustworthiness of the subsystems and components itself is subject to uncertainty [1-4]. Besides, evaluating the models for the trustworthiness of complex systems are needed to be capable of modeling the uncertainty and also calculate and express the degree of uncertainty associated to the derived trustworthiness of the overall system. Trust is interrelated with people's everyday life. When people want to do some new things like selling or buying things from one to other, finding new materials from different source or some other things, concept of trust then arise. In the field of computer science and virtual world of modern technologies like virtual organizations, mobile or P2P networks and E-commerce [5-7], trust is very important. One to one conversion, data sharing is fully dependent on this trust on something. Different trust models have been developed now a day for serving this trustworthiness in the virtual communication world which is seen mostly worked on uncertainty.

Following the day by day improvements of the internet of services, the future internet based on Cloud computing IT systems will become highly distributed, dynamically composed and will be hosted and managed by multiple parties. But it is sorry to say at present people, enterprises, officials, organizations and corporate farms are still hesitating and feeling less of security and safety to move to the Cloud [8-10]. The reasons behind this are missing transparency, security concerns. So, both the users and providers and accreditation authorities are interested in evaluating the trustworthiness of a service, infrastructure or platform.

It is evident that the evaluation of the trustworthiness of complex systems is one of the major challenges in current IT research. Different trust models are now present in the world, which are dependent on uncertainty. [11-15] A new proposed model for solving this problem is Certain Trust Model (CTM) [1] which is used to calculate the trust of a system depending on recommendation of some experts, means on some certain values.

But, this model has some limitations. It has failed to apply fuzzy logic, probabilistic logic. This model is developed on the basis of human understanding. But for machine understanding and taking decisions, fuzzy logic is much helpful [16]. Again, it is much helpful for human beings to understand any situation and taking different type of decisions about something with the help of fuzzy logic [17-19] rather than other logics. The goal of our work is to extend representational model of CTM with the help of fuzzy logic, probabilistic logic so that the model can overcome its limitations. We have designed two new parameters for this purpose. These parameters are developed based on certain trust logic [2], which is dependent on CTM. The representational model of CTM will become friendlier to the users and the developers and make it more appropriate than the previous version of CTM.

Here, in this paper, section 2 describes the describes related work of our proposed work; section 3 describes briefly the model on which we work, section 4 describes our proposed model and implementation process, section 5 shows some case studies of our work, section 6 shows experimental results of our model which we had run in the lab.

## II. RELATED WORK

Several number of ways are there for modeling (un-)certainty of trust values in the field of trust modeling in Cloud computing and internet based marketing sys-tem.[12, 13, 32] But, these models have less capability to derive trustworthiness of a system which are based on knowledge about its components and subsystems. The main challenge of these models is to find out good models for deriving trust using three ways:-

1. Trust from direct experience of a user,

2. Recommendations from third parties, '

3. Additional information.

These models help to save from robustness to attacks, e.g. misleading recommendations, Sybil attacks, etc. [20].

Fuzzy logic was used to provide trust in Cloud computing. Different types of attacks and trust models in service oriented systems, distributed system and so on are designed based on fuzzy logic system [21-24]. But it models different type of uncertainty known as linguistical uncertainty or fuzziness [17]. In paper [18], a very good model for E-commerce, which is based on fuzzy logic, is presented. But, this model also works with uncertain behavior. Belief theory such as Dempster-Shafer theory was used to provide trust in Cloud computing [20]. But the main drawback of this model is that the parameters for belief, disbelief and uncertainty are dependent on each other. It is possible to model uncertainty using Bayesian probabilities[25] which lead to probability density functions e.g., Beta probability density function. It is also possible to apply the propositional standard operators to probability density functions. But this leads to complex mathematical operations and multi-dimensional distributions which are also hard to interpret and to visualize. An enhanced model recently being developed for using in Cloud computing is known as Certain Trust. This model evaluates propositional logic terms that are subject to uncertainty using the propositional logic operators AND, OR and NOT[1-4].

## III. CERTAIN TRUST MODEL

Certain Trust Model was constructed for modeling those probabilities, which are subject to uncertainty. This model was designed with a goal of evidence based trust model. Moreover, it has a graphical, intuitively interpretable interface [1] which helps the users to understand the model (Fig 1). The representational model focuses on two crucial issues

*a) How trust can be derived from evidence considering context-dependent parameters?*

*b) How trust can be represented to software agents and human users?*

For the first one, a relationship between trust and evidence is needed. For this, they had chosen a Bayesian approach. It is because it provides means for deriving a subjective probability from collected evidence and prior information [1]. At developing a representation of trust, it is necessary to consider to whom trust is represented. It is easy for a software component or a software agent to handle mathematical representations of trust. For it, Bayesian representation of trust is appropriate. The computational model of Certain Trust proposes a new approach for aggregating direct evidence and recommendations. In general, recommendations are collected to increase the amount of information available about the candidates in order to improve the estimate of their trustworthiness. This recommendation system needs to be integrated carefully for the candidates and for the users and owner of cloud servers. This is called robust integration of recommendations. In order to improve the estimate of the trustworthiness of the candidates, it is needed to develop recommendation system carefully. This is called robust integration of recommendations [1, 2].

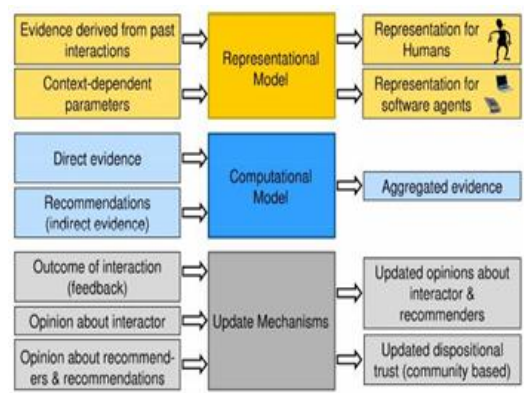

Figure 1. Block diagram of Trust models

Three parameters used in certain logic: average rating t, certainty c, initial expectation f. The average rating t indicates the degree to which past observations support the truth of the proposition. The certainty c indicates the degree to which the average rating is assumed to be representative for the future. The initial expectation f expresses the assumption about the truth of a proposition in absence of evidence [1-4].

The equations for these parameters are given below:-

Equation for average rating, $t = \begin{cases} 0.5 & if\ r + s = 0 \\ r/(r+s) & else \end{cases}$ (1)

Here, r represents number of positive evidence and s represents number of negative evidence defined by the users or third person review system.

Equation for certainty,

$$c = \frac{N.(r+s)}{2.w.(N-(r+s))+N.(r+s)}$$ (2)

Here, w represents dispositional trust which influences how quickly the final trust value of an entity shifts from base trust value to the relative frequency of positive outcomes and N represents the maximum number of evidence for modeling trust. Using these parameters the expectation value of an opinion $E(t,c,f)$ can be defined as follows:

$$E(t,c,f) = t*c + (1-c)*f$$ (3)

The parameters for an opinion o = (t, c, f) can be assessed in the following two ways: direct access and Indirect access. Certain Trust evaluates the logical operators of propositional logic that is AND, OR and NOT. In this model these operators are defined in a way that they are compliant with the evaluation of propositional logic terms in the standard probabilistic approach. However, when combining opinions, those operators will especially take care of the (un)certainty that is assigned to its input parameters and reflect this (un)certainty in the result. The definitions of the operators as defined in the CTM are given in the table 1.

TABLE I. DEFINITION OF OPERATORS

| | |
|---|---|
| OR | $c_{A \vee B} = c_A + c_B - c_A c_B - \frac{c_A(1-c_B)f_B(1-t_A)+(1-c_A)c_B f_A(1-t_B)}{f_A+f_B-f_A f_B}$ <br> $t_{A \vee B} = \begin{cases} \frac{1}{c_{A \vee B}}(c_A t_A + c_B t_B - c_A c_B t_A t_B) & if\ c_{A \vee B} \neq 0 \\ 0.5 & else \end{cases}$ <br> $f_{A \vee B} = f_A + f_B - f_A f_B$ |
| AND | $c_{A \wedge B} = c_A + c_B - c_A c_B - \frac{(1-c_A)c_B(1-f_A)t_B + c_A(1-c_B)(1-f_B)t_A}{1-f_A f_B}$ <br> $t_{A \wedge B} = \begin{cases} \frac{1}{c_{A \wedge B}}(c_A c_B t_A t_B + \frac{c_A(1-c_B)(1-f_A)f_B t_A + (1-c_A)c_B f_A(1-f_B)t_B}{1-f_A f_B}) & if\ c_{A \wedge B} \neq 0 \\ 0.5 & else \end{cases}$ <br> $f_{A \wedge B} = f_A f_B$ |
| NOT | $t_{\neg A} = 1 - t_A,\quad c_{\neg A} = 1 - c_A\ \ and\ \ f_{\neg A} = 1 - f_A$ |

IV. PROPOSED APPROACH AND USED CASES

For developing and exercising with our work, we have used a scenario from the field of cloud computing [1]. We have assumed that we are working to evaluate the trustworthiness of an organization or a simple office. We have worked mainly for the field of trade and business web pages. It has helped us to calculate the trust of the whole cloud computing system and also helped to make a system trustworthy to the user.

In this test system (figure 2), the server S directly relies on two subsystems or servers, S1 and S2. Subsystem S1 consist of two servers (A1 and A2), where at least one of the servers has to be available. Similarly, subsystem S2 is composed of two redundant data bases servers (only one need to be available).

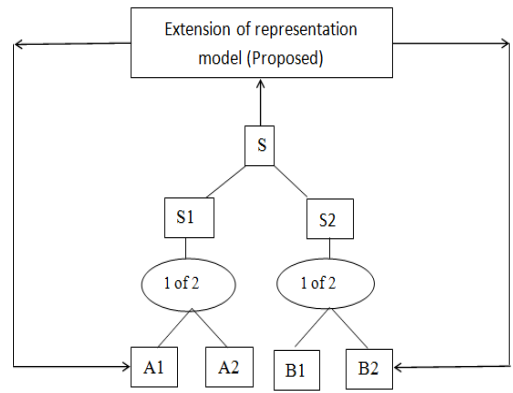

Figure 2. Assumed Cloud Architecture

Based on the description above and getting the information about the trust values of the atomic components, the evaluation of the trustworthiness of the complete system in the context of availability, can be carried out by evaluating the following propositional logic term:

$$(A1 \vee A2) \wedge (B1 \vee B2)$$ (4)

According to the certain trust, the average rating t indicates the degree to which past observations support the truth of the proposition. t = 0 implies that there is only evidence contradicting the proposition. t = 1 implies that there is only evidence supporting the proposition. According to us, this average rating can be scaled according to the wish of the developer. E.g. One developer wants to scale the rating of his product from 1 to 5, means 1 is the lowest rate of the product and 5 is the highest rate. This is one type of **Rating Based Trust Model** [33] approach. So, for him, this average rating can be scaled between 1 and 5. From the rate of the product, one user or buyer can take his decision about the product and keep him safe from being betrayed. Again, another parameter, named certainty c, indicates the degree to which the average rating is assumed to be representative for the future. The higher the certainty, the higher is the influence of the average rating on the expectation value in relation to the initial expectation. When the maximum level of certainty is reached, the average rating is assumed to be representative for the future outcomes. c = 0 implies that there is no evidence available. c = 1 implies that the collected evidence is considered to be representative [1]. Here, the certainty c not only takes the value of 0 or 1 but also the values between 0 and 1. The scaling of this parameter depends on the interest of the developer or the manager of the office or industry. It mainly depends on the number of evidence. The last parameter of certain logic is initial expectation f. The initial expectation f expresses the assumption about the truth of a proposition in absence of evidence. It is helpful for the new owner or developer of a product to express his expectation about the product's service to humans. [1, 2, 3].

After taking the output from the CTM, we have applied *fuzzy logic* on it. It is an extension of the representational model of CTM. After getting value of c, t and f at the system top position S, our model starts working. We have plotted this in basic fuzzy logic system. After calculating, the result will be send to the lower level users, means to the lower level servers,

PCs and system. According to our figure 2, these are $A_1$, $A_2$, $B_1$, $B_2$.

With the help of these parameters and operators derived from certain trust, we have defined two new parameters, Trust T and behavioral probability, P. Trust T is calculated from certainty c and average rating t. the equation is:

$$\text{Trust, T} = \frac{c*t}{Highscalingvalueofrating} * 100\% \quad (5)$$

Here, High scaling value of rating means the upper value of the range of rating.

Calculating T, we have applied FAM rule of fuzzy logic for creating a relation between certainty c and average rating t. Trust T represents this relation in percentage such a way that the quality of the product can easily be understood.

Another parameter, behavioral probability, P, represents how the present behavior of the system varies from its initially expected value and it is proposed with the help of probabilistic logic [30].

It may be less, equal or higher than the initial expectation given by the system developer or the manager of the office. The equation for P is:

$$\text{Behavioral probability, P} = \frac{(T)-f}{f} * 100\% \quad (6)$$

If T<f, lower probability to show expected behavior

If T>f, higher probability to show expected behavior

If T=f, balanced with the expected behavior

From the equation of P, values with two type magnitude have been found. If it is negative, then it is assumed that it will behave lower than the expected. If it is positive, then higher behavior will be shown by the system than the expected behavior of this system, which is defined by the developer or someone related to the system.

We can see following values for behavioral probability, P.

TABLE II.   BEHAVIORAL PROBABILITY FOR DIFFERENT RANGES

| Trust Ranges | Calculated P | Comment |
|---|---|---|
| 1-20% | 98%-60% | Lowest Behavior |
| 21-40% | 58%-20% | Lower Behavior |
| 41-49% | 18%-2% | Low Behavior |
| 50% | 0 | Balanced |
| 51-60% | 2%-20% | High Behavior |
| 61-80% | 22%-60% | Higher Behavior |
| 81-100% | 62%-100% | Highest Behavior |
| >100% | 100% | Highest Behavior |

Where, initial expectation f is assumed to be 0.5; means showing 50% of accurate behavior of the system or the product. We have run this whole system in our lab and have got related results discussed above.

Observing the value of P, one can easily understand whether the system can fulfill his expectation or not according to the expectation of the developer about the whole cloud computing system or the trading product. These two parameters help both the developer and user.

*A. Fuzzification and Defuzzification*

A fuzzy logic system (FLS) can be defined as the nonlinear mapping of an input data set to a scalar output data [25-29]. A FLS consists of four main parts: fuzzier, rules, inference engine, and defuzzier. Fuzzy logic consists of following components are: -

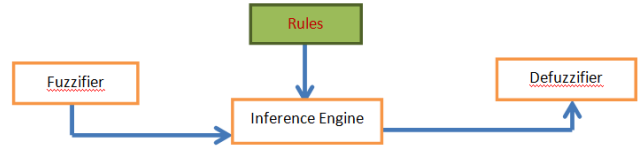

Figure 3. Basic components of fuzzy model

The process of fuzzy logic maintains the following steps: Firstly, a crisp set of input data are gathered and converted to a fuzzy set using fuzzy linguistic variables, fuzzy linguistic terms and membership functions. This step is known as fuzzification. Afterwards, an inference is made based on a set of rules. Lastly, the resulting fuzzy output is mapped to a crisp output using the membership functions, in the defuzzification step. Fuzzification is a process where inputs are a set of fuzzy inputs and the output is crisp values.

*B. Fuzzy Inputs*

According to Gaussian, the membership function for fuzzy input sets depends on two types of parameter, standard deviation **σ** and mean **c.** The equation for membership function is:-

$$f(x; \sigma; c) = \exp\left(\frac{-(x-c)^2}{2\sigma^2}\right) \quad (7)$$

In designing fuzzy inference system, it is easy to understand that membership functions are associated with term sets, which normally appears in the antecedent or consequent of rules. We have divided parameter certainty c into five categories according to its values. They are:-

TABLE III.   RANGES OF CERTAINTY

| Class Name | Certainty Range Value | Symbols |
|---|---|---|
| Very Low | 0.0-0.2 | VLc |
| Low | 0.1-0.4 | Lc |
| Average | 0.3-0.7 | Avg.c |
| High | 0.6-0.9 | Hc |
| Very High | 0.8-1.0 | VHc |

Following the same way, we have divided parameter average rating t into five categories according to its values in table IV:-

Though we classify the parameters value according to the ranges described above, it can be varied from persons to persons. For this, we take helps from fuzzy logic.

TABLE IV. RANGES OF AVERAGE RATING

| Class Name | Avg. Rating Range Value | Symbols |
|---|---|---|
| Very Low | 1.0-2.0 | VLt |
| Low | 1.5-3.0 | Lt |
| Average | 2.0-4.0 | Avg.t |
| High | 3.0-4.5 | Ht |
| Very High | 4.25-5.0 | VHt |

Following the Gaussian membership function equation, we have got the figures stated below:-

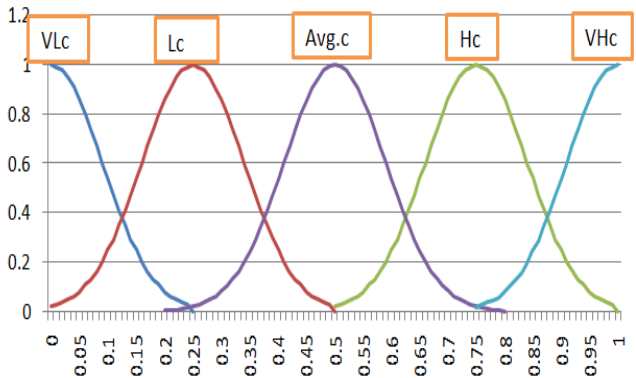

Figure 4. Membership Functions for certainty

Here X- axis represents the certainty deviation.

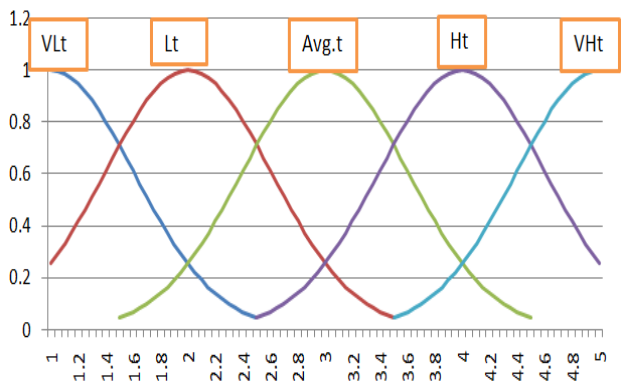

Figure 5. Membership Functions for average rating

### C. Inference Rules

Fuzzy inference is the process of formulating the mapping from a given input to an output using fuzzy logic. The mapping then provides a basis from which decisions can be made, or patterns discerned. There are two concepts of fuzzy logic systems [27]. They are: - linguistic variables and fuzzy if then else rule. The linguistic variables' values are words and sentences where if then else rule has two parts; antecedents and consequent parts which contain propositions of linguistic variables. Numerical values of inputs $x_i \epsilon U_i (i=1,2….n)$ are fuzzified into linguistic values, $F_1, F_2…..F_n$. Here $F_i$ denotes the universe of discourse $U = U_1 * U_2 * …….. * U_n$. The output linguistic variables are $G_1, G_2, ……, G_n$. The if-then-else rule can be defined as: -

$R^{(j)}$: IF $x_i \epsilon F_1^j$ and ……and $x_n \epsilon F_n^j$ THEN $y \epsilon G^j$. (8)

Where, j = 1,2,….., M. M is the number of rules.

According to rules discussed above, we have proceeded for our proposed model. There are 25 fuzzy rules in our extension model. They are (R represents rule):-

**R1:-** If certainty is very low and average rating is very low, then trust is very low.

**R2:-** If certainty is low and average rating is very low, then trust is very low.

**R3:-** If certainty is average and average rating is very low, then trust is very low.

**R4:-** If certainty is high and average rating is very low, then trust is very low.

**R5:-** If certainty is very high and average rating is very low, then trust is very low.

**R6:-** If certainty is very low and average rating is low, then trust is very low.

**R7:-** If certainty is low and average rating is low, then trust is low.

**R8:-** If certainty is average and average rating is low, then trust is low.

**R9:-** If certainty is high and average rating is low, then trust is average.

**R10:-** If certainty is very high and average rating is low, then trust is average.

**R11:-** If certainty is very low and average rating is average, then trust is very low.

**R12:-** If certainty is low and average rating is average, then trust is low.

**R13:-** If certainty is average and average rating is average, then trust is average.

**R14:-** If certainty is high and average rating is average, then trust is average.

**R15:-** If certainty is very high and average rating is average, then trust is high.

**R16:-** If certainty is very low and average rating is high, then trust is very low.

**R17:-** If certainty is low and average rating is high, then trust is low.

**R18:-** If certainty is average and average rating is high, then trust is average.

**R19:-** If certainty is high and average rating is high, then trust is high.

**R20:-** If certainty is very high and average rating is high, then trust is high.

**R21:-** If certainty is very low and average rating is very high, then trust is very low.

**R22:-** If certainty is low and average rating is very high, then trust is low.

**R23:-** If certainty is average and average rating is very high, then trust is average.

**R24:-** **If certainty is high and average rating is very high, then trust is high.**

**R25:-** **If certainty is very high and average rating is very high, then trust is very high.**

According to these inference rules stated above, we have got the fuzzy input sets shown in figure 4 and figure 5. From that, we have got figure 6 output crisp values.

### D. Fuzzy Outputs

From the input fuzzy sets described above, passing those fuzzy sets through inference rules and fuzzy base rules, we get crisp values for our new parameter trust T. Plotting those values according to Gaussian membership function equation we have got the figure… for Trust T parameter. It can also be classified into five categories after finding out and plotting:-

TABLE V. RANGES OF OUTPUT TRUST

| Class Name | Trust Range Value | Symbols |
|---|---|---|
| Very Low | 0%-20% | VLT |
| Low | 10%-40% | LT |
| Average | 30%-70% | Avg.T |
| High | 60-90% | HT |
| Very High | 80%-100% | VHT |

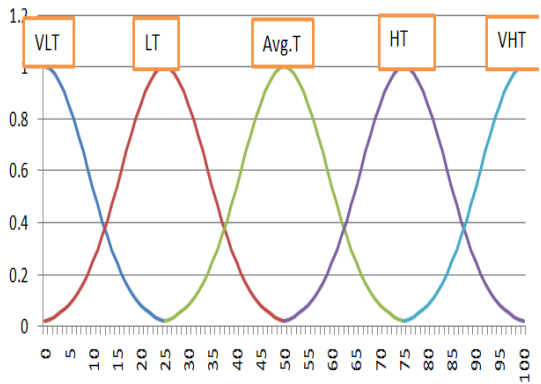

Figure 6. Membership Functions for Output Trust

Here X-axis represents the trust values.

### E. Defuzzification

The input for the defuzzification process is a fuzzy set and the output of defuzzification process is a crisp value obtained by using some defuzzification method such as centroid, height and maximum. Among them, centroid defuzzification is used mostly. We use the following equation for applying Defuzzification method:-

$$y' = \frac{\sum_{i=1}^{n} y_i \mu(y_i)}{\sum_{i=1}^{n} \mu(y_i)} \quad (9)$$

### F. Applying Implication Method

The prerequisite for applying implication to any fuzzy set is finding out rule's weight.

We have found out the weights in figure 6 Membership functions output is de-fined as the weights for every rules.

The input for the implication process is a single number given by the antecedent, and the output is a fuzzy set. Implication is implemented for each rule.

Let, one of the rules is:-

"If certainty is high and average rating is aver-age, then trust is average."

Let, the value for certainty is C=0.7 and value for average rating is t = 3.0. Now, according to the implication method, we get the following output:-

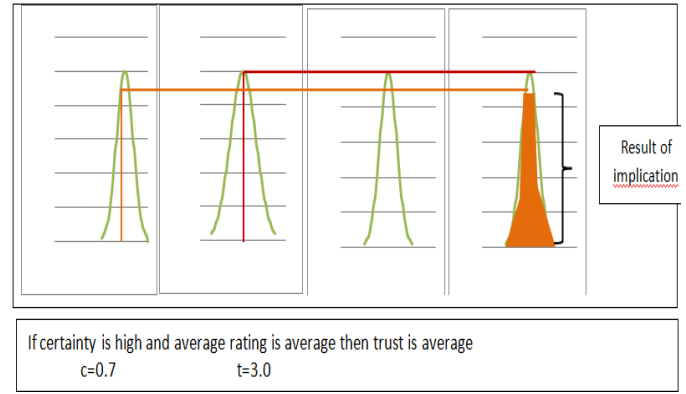

Figure 7. Implication for R14

### G. Aggregate all Outputs

Aggregation is the process by which the fuzzy sets that represent the outputs of each rule are combined into a single fuzzy set. Aggregation only occurs once for each output variable.

It is the second last phase of defuzzification. The input of the aggregation process is the list of truncated output functions returned by the implication process for each rule. The output of the aggregation process is one fuzzy set for each output variable. In this proposed model, the fuzzy input set is certainty set and average rating set and the output fuzzy set is trust set.

### H. Defuzzification Results

Using defuzzification equation no (9), we can get the defuzzified output. According to it, the defuzzified output is:-

$$y' = \frac{1783.81}{32.851} = 54.3 \quad (10)$$

And it continues.

### I. Mapping Surface

In this map, we plot certainty, c and average rating, t and Trust, T. after plotting this, we get the following surface.

## V. CASE STUDIES

In this section, we are going to show the impact of newly defined operators over the operators of CTM. Following this, we will show the impact of our model when it is applied is a private server or cloud. Two cases are described here, case 1 for multiple servers and case 2 for single server.

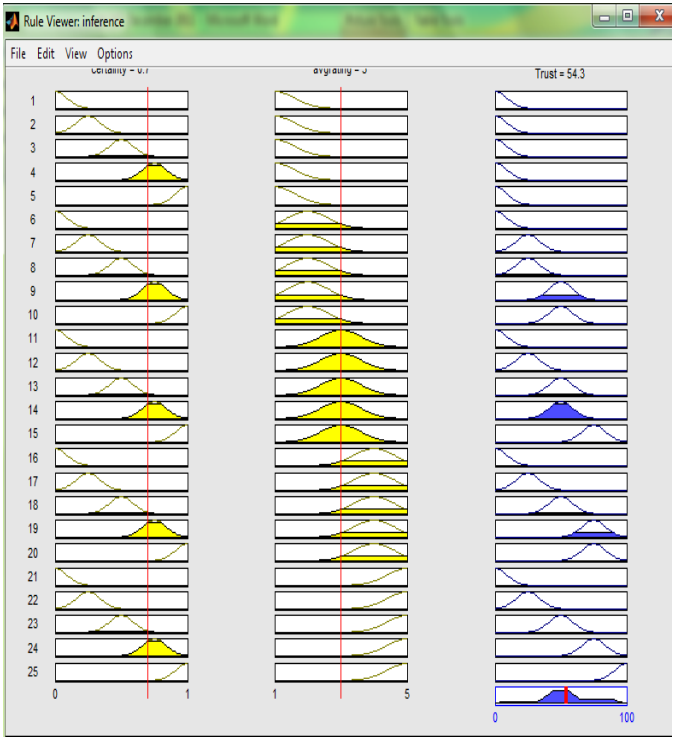

Figure 8. Sequential Process of Aggregating all outputs

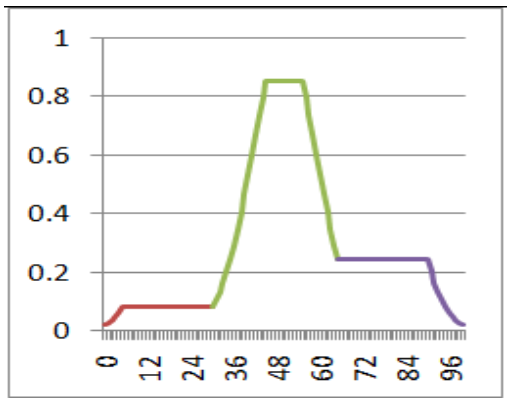

Figure 9. Defuzzified Output After Aggregation

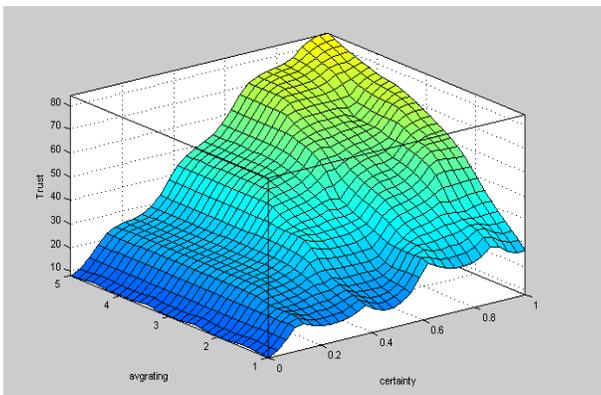

Figure 10. Defuzzified Output After Aggregation

**Case 1:** According to CTM's operators defined in equation 3, 4, and 5, we know that, the input for this model is r, s, f and w. Let, the input values are r=5, s=2, f=0.5 and w=1. Here, no of evidences are N=7.Then, the output values are:- average rating t = 0.714 and c=0.724. and then, E = 0.65. Now, for mapping it to our proposed model, we need to modify t. Here,

$$t' = t*\text{scale of rating} \qquad (11)$$

Usually, the scale of rating is 5. Now, the new average rating is t = 0.714*5 = 3.57. Then, the value of parameter Trust, T = ((3.57*0.724)/5)*100 = 51.69%. From fig…, we see that, it is an average situation of Trust. As the range of trust varies from person to person, one can consider it under high trust region. From the result of E and T, we can see that, it is easier to understand the condition of system or server much better that the past. As it is represented in percentage form, the user can easily understand the evaluated value of trust. Now, the value of second parameter, behavioral probability, p = 3.39% and because of T>f, the system is now in the condition of showing 3.39% higher behavior than the initial expectation. As it is in higher condition, so, the hosting partner can easily take the decision to host in this server/system. This parameter will also be useful for the developers so that they can easily understand the present condition of the system.

Now, for the system shown in figure 2 and with the help of equations given in Table No 1, we have seen the following situation: - (considering the above described values as system A)

TABLE VI. OUTPUT OF CTM AND PROPOSED MODEL

| System | Let the values | Output for Certain Trust Model | Output for our proposed model |
|---|---|---|---|
| $A_1$*** | $t_{A1}$=0.714, $c_{A1}$=0.724, | $E_{A1}$=0.65 | $T_{A1}$ = 51.69,M, $P_{A1}$= 3.39 higher |
| $A_2$*** | $t_{A2}$= 0.459, $c_{A2}$= 0.806, | $E_{A2}$ =0.467 | $T_{A2}$ = 37, M/L, $P_{A2}$ = 26 lower |
| $B_1$*** | $t_{B1}$=0.604, $c_{B1}$= 0.786, | $E_{B1}$= 0.582 | $T_{B1}$ = 47.47, M, $P_{B1}$ = 5 lower |
| $B_2$*** | $t_{B2}$=0.867, $c_{B2}$= 0.648, | $E_{B2}$ = 0.74 | $T_{B2}$ = 56.18, M, $P_{B2}$ = 12.36 higher |
| $S_1$ | $t_{S1}$=0.829,$c_{S1}$=0.839, $f_{S1}$=0.75, | $E_{S1}$=0.82 | $T_{S1}$ = 69.55, H, $P_{S1}$ = 7.26 lower |
| $S_2$ | $t_{S2}$=0.892, $c_{S2}$=0.863, $f_{S2}$=0.75, | $E_{S2}$=0.87 | $T_{S2}$ = 77, H, $P_{S2}$ = 2.67 higher |
| S | $t_s$=0.736, $c_s$=0.853, $f_s$=0.5625, | $E_s$=0.753 | $T_S$ = 62.78, M, $P_A$ = 11.61 higher |

*** *for all cases, f = 0.5.*

**Case 2:** Our proposed model can also be applied for a website hosted in single server and calculating trust for a product in online transaction. Like [31], each of its portion e.g. existence, policy, fulfillment and affiliation can be represented in certainty and average rating format. And from that, we can easily calculate a website's or product trust.

Let, one person makes a target to buy a product through online. For this, that person must want to check the trustworthiness of that website from where he is going to start his business, peoples review and the quality of the product and people review of that product. One can also want to check the number of sold of that product. With our proposed model, it can easily be designed. The person, who wants to buy a product through online, can get information about it from the closer persons or someone who has an experience about the product or about the website. If it is, then that put 1 for certainty, c. That means, c=1.0.

If he is fully new to that website or product, then c=0. At the time of buying that product from website, he can give a rate to that website from different axis, such that: - served information about the product, easiness, service, privacy information, physical existence of that product, registration process, whether the website deals with updated product or not and so on.

Then, taking the average of the rating of these factors, we can get the average rating of that website. Let, average rating, t =3.75. Then, according to our model, if, c = 1 for that person, then, trust, T = 75%. It is given by that person only. Now, according to the developers of that website, initial expectation f and ranging of certainty is defined. If the developers want to take minimum 20 people's certainty, means want to take response of c=1 from at least 20 people for their system's or website's accuracy, then they can scale the certainty parameter c in following way: -

TABLE VII. Scaling of certainty for 20 people

| Range of people | 0<n<=4 | 4<n<=8 | 8<n<=12 | 12<n<=16 | 16<n<=20 |
|---|---|---|---|---|---|
| Certainty, c | 0.2 | 0.4 | 0.6 | 0.8 | 1.0 |

Now, for this, we can get the values for T shown in table VIII; applying the above table of values for 20 people of the organization we have explained earlier:-

TABLE VIII. FAM for representing trust (20 people)

| c \ t | 1 | 2 | 3 | 4 | 5 |
|---|---|---|---|---|---|
| 0.0 | N | N | N | N | N |
| 0.2 | VL | VL | VL | VL | VL |
| 0.4 | VL | VL | L | L | L |
| 0.6 | VL | L | L | M | M |
| 0.8 | VL | L | M | H | H |
| 1.0 | VL | L | M | H | VH |

Here, for average rating t, we have shown only scaled values. These values can also be accepted for fractional values of average rating. The rule matrix will be needed to design in the same way for those fractional average ratings.

If, the developers have a good confidence about their hosted website and after checking different mandatory requirements for that website, e.g. ensuring security level, payment method, delivery system, etc, they give the value of initial expectation, f = 0.5, then, behavioral probability, p = 50% upper. This means that, the present condition of this website is 50% upper than the initial expectation of the developers.

Let, we take 100 people/experts evidence or transaction as measurement limitation for measuring trust for experimental purpose. From that, using the basis of the CTM we have got different values for certainty c, average rating t and initial expectation f.

Applying fuzzy logic according to the table II on the experimental values we have got the values for T shown in table IX, which specify the trustiness of the system.

With the help of the FAM shown above the behavioral probability of the system can easily be calculated. With the help of T and P, one trader or user can easily take decision about the trustworthiness of the system, especially for cloud computing and the trading product. For calculating a single product's trust or rating, we have tested with 20 people's comment and experience. Secondly, we have experimented with our test system shown in (figure 2) with 100 people's experience and evidences. It seems easier for the users and developers to understand the trust and behavioral probability of the system and trust value of a single product system.

TABLE IX. FAM FOR REPRESENTING TRUST (100 PEOPLE)

| c \ t | 1.0 | 1.5 | 2.0 | 2.5 | 3.0 | 3.5 | 4.0 | 4.5 | 5.0 |
|---|---|---|---|---|---|---|---|---|---|
| 0.0 | N | N | N | N | N | N | N | N | N |
| 0.1 | VL | VL | VL | VL | VL | VL | VL | VL | VL |
| 0.2 | VL | VL | VL | VL | VL | VL | VL | VL | VL |
| 0.3 | VL | VL | VL | VL | VL | L | L | L | L |
| 0.4 | VL | VL | VL | VL | L | L | L | L | L |
| 0.5 | VL | VL | VL | L | L | L | L | M | M |
| 0.6 | VL | VL | L | L | L | M | M | M | M |
| 0.7 | VL | L | L | L | M | M | M | H | H |
| 0.8 | VL | L | L | L | M | M | H | H | H |
| 0.9 | VL | L | L | M | M | H | H | VH | VH |
| 1.0 | VL | L | L | M | M | H | H | VH | VH |

Our proposed representational model consists of two parameters which give us benefits according to the following points of view. These comparisons are held on the basis of paper [1], [2] and [31]:-

TABLE X. Comparison Between Models

|  | Original Certain Trust Model | Ecommerce fuzzy trust model[31] | Our proposed model |
|---|---|---|---|
| Fuzziness | No | Yes | Yes |
| Behavioral parameters | No, but one can assume | No | Yes |
| Certainty | Yes | No | Yes |
| Probabilistic Logic | Probability theory | No | Combination of Probability Theory and logic |
| Advantage | For Human interaction, its HTI is easy to understand | Useful for people | Useful for people and for machine. |

## VI. CONCLUSIONS

In this paper, we have proposed a new extension of representational model of certain trust for the evaluation of propositional logic terms, probability and fuzziness under uncertainty. It develops the representational model of the certain trust logic. Our proposed model is more expressive and useful than certain logic because it works both for machine and human beings. The parameters of the proposed model directly show how much the system can be trusted and it can be applied not only in small systems but also large systems; especially in cloud computing field. Again, it represents the present condition of the product, website and also for the system.

We have some new idea to imply in our proposed model in future. Firstly, we want to apply evolutionary algorithm with this model to optimize and design the rules. We want to apply price comparison as a parameter for a product in our model for ensuring accurate trust measuring model for a normal e-commerce website. Secondly, more development of behavioral probability parameter so that it can directly prohibit different types of security breaking questions like Sybil attack, false rating, etc is our fourth wish. At present, this parameter works indirectly with security options. Last of all, we want to establish a newer trust model with a combination of certainty, fuzzy logic, evolutionary algorithm and so on for ubiquitous computing system like cloud computing, distributed computing, etc.

ACKNOWLEDGMENT

The authors wish to thank Sebastian Ries, Sheikh Mahbub Habib, Max Mühlhäuser and Vijay Varadharajan for their renownable work and thinking.

REFERENCES

[1] Ries, S.: Trust in Ubiquitous Computing. PhD thesis, Technische University at Darmstadt,pp: 1-192, 2009
[2] Sebastian Ries, Sheikh Mahbub Habib, Max Mühlhäuser and Vijay Varadharajan: Certain Logic: A Logic for Modeling Trust and Uncertainty, Technical report, April 6th, 2011
[3] Sebastian Ries, Sheikh Mahbub Habib, Max Mühlhäuser and Vijay Varadharajan: Certain Logic: A Logic for Modeling Trust and Uncertainty (Short Paper), June, 2011.
[4] Sebastian Ries, "Certain Trust: A Trust model for Users and Agents ", ACM, March 2007, pp. 1599-1604.
[5] Adoption, Approaches & Attitudes "The Future of Cloud Computing in the Public and Private Sectors", A Global Cloud Computing Study, JUNE 2011.
[6] Buchegger, S., Le Boudec, J.Y.A Robust Reputation System for Peer-to-Peerand Mobile Ad-hoc Networks. In: P2PEcon 2004.
[7] Junfeng TIAN, Chao LI, Xuemin HE, Rui TIAN, "A Trust Model Based on The Multinomial Subjective logic for P2P Network", International J. communications, Network and Systems, 2009, pp. 546-554
[8] ENISA: An SME perspective on cloud computing - survey. Technical report, ENISA (2009)
[9] Chow,R.,Golle,P.,Jakobsson,M.,Shi,E.,Staddon,J.,Masuoka,R.,Molina,J. : Controlling data in the cloud: outsourcing computation without outsourcing control. In: Proceedings of the 2009 ACM workshop on Cloud computing security. CCSW '09, ACM (2009) 85–90
[10] Armbrust, M., Fox, A., Griffith, R., Joseph, A.D., Katz, R.H., Konwinski, A., Lee, G., Patterson, D.A., Rabkin, A., Stoica, I., Zaharia, M.: Above the clouds: A berkeley view of cloud computing. University of California, Berkeley (Feb 2009)
[11] D.W. Manchala, "E-Commerce Trust Metrics and Models", IEEE Internet Computing, March-April 2000, pp.36-44.
[12] Mohammed Alhamad, Tharam Dillon, and Elizabeth Chang A Trust-Evaluation Metric for Cloud applications *International Journal of Machine Learning and Computing, Vol. 1, No. 4, October 2011.*
[13] Teacy, W., et al.: Travos: Trust and reputation in the context of inaccurate informationsources. Aut.Agentsand Multi-Agent Systems, vol.12, no.2, pp. 183-198, 2006.
[14] J_sang, A., McAnally, D.: Multiplication andco-multiplication of beliefs. International Journal of Approximate Reasoning vol. 38 no. 1,pp. 19-51, 2005.
[15] Jøsang, A.: A logic for uncertain probabilities. International Journal of Uncertainty, Fuzziness and Knowledge-Based Systems9(3) (2001) 279–212
[16] TJ Ross, "Fuzzy Logic with engineering applications", Wiley Online Library, 2005.
[17] Lotfi A. Zadeh, "Toward a theory of fuzzy information granulation and its centrality in human reasoning and fuzzy logic", Science Direct, 13 May 1998.
[18] Vojislav Kecman, "Learning and Soft computing", Google online library, 2001.
[19] Y. Lin, W.J. Zhang, C. Wu, G. Yang, J. Dy, "A fuzzy logics clustering approach to computing human attention allocation using eyegaze movement cue", International Journal of Human-Computer Studies, Volume 67, Issue 5, May 2009, Pages 455-463
[20] Kerr, R., Cohen, R.: Smart cheaters do prosper: defeating trust and reputation systems. In: AAMAS '09: Proceedings of The 8th International Conference on Autonomous Agents and Multiagent Systems, (2009) 993–1000
[21] Cristiano Castelfranchi, Rino Falcone, Giovanni Pezzulo, "Trust in information sources as a source for Trust: A fuzzy Approach ", ACM, 2003, pp- 89-96
[22] Jøsang, A., Ismail, R., Boyd, C.: A survey of trust and reputation systems for online service Provision. Decision Support Systems 43(2) (2007) 618–644
[23] Elizabeth J.Chang, Farookh Khadeer Hussain, Tharam S. Dillon "Fuzzy Nature of Trust and Dynamic Trust Modeling in Service Oriented Environments", ACM, SWS'05, November 11, 2005, Fairfax, Virginia, USA.
[24] Florian Skopik, Daniel Schall, Schahram Dustdar, "Trustworthy Interaction Balancing in Mixed Service-oriented Systems", ACM 2010, SAC'10 March 22-26, 2010, Sierre, Switzerland
[25] L. Zadeh. "Fuzzy sets", Journal of Information and Control, 8:338—353, 1965
[26] Fuzzy Logic Fundamentals, chapter 3, pp-61-84, March, 2001
[27] Shrija Rajbhandari, Omer F. Rana and Ian Wootten, School of Computer Science, Cardiff University, Cardiff, U.K," A Fuzzy Model for Calculating Workflow Trust using Provenance Data", ACM, 2008, Baton Rouge, USA.

[28] Martin Chun-Sheng Cheng, Dynamical near optimal training for interval type-2 fuzzy neural network with genetic algorithm, thesis paper, pp-1-30, 2003

[29] .N.N. Karnik, J.M. Mendel, and Q. Liang, "Type-2 fuzzy logic system", IEEE Trans. on Fuzzy Syst., vol. 7 no. 6, pp. 643-658, Dec 1999.

[30] Nils J. Nilson, "Probabilistic Logic*", Artificial Intelligence 28, Elsevier Science Publishers B.V., pp – 71-87

[31] Samia, Nefti, FaridMeziane, KhairudinKasiran, A Fuzzy Trust Model for E-Commerce, 2004

[32] Ries, S., Heinemann, A.: Analyzing the robustness of CertainTrust. In: 2nd JointTrust and PST Conf. on Privacy, Trust Management and Security, pp. 51-67, 2008.

[33] Li Xiong, Ling Liu, A Reputation-Based Trust Model for Peer-to-Peer e Commerce Communities, 2003.

AUTHORS PROFILE

**Kawser Wazed Nafi** is now working as Lecturer in Computer Science and Engineering department of Stamford University, Bangladesh. He completed his graduation from Khulna University of Engineering and technology in Computer Science and Engineering Department. He is very ambitious and wanted to work with artificial intelligence, trust models and security in networking systems. His favorite research field is cloud computing, distributed computing, parallel computing, network security and so on.

**Tonny Shekha Kar** is now working as Lecturer in Computer Science and Engineering department of Stamford University, Bangladesh. She completed her graduation from Khulna University of Engineering and Technology in Computer Science and Engineering department. Her research interest is cloud computing, security and trust issues in networking systems.

**MD. Amjad Hossain** completed his graduation from Computer Science and Engineering department of Khulna University of Engineering and Technology and became lecturer of that department. He is now completing his PhD in Kent State University, USA. His main research interests are Cloud Computing, Quantum Computing, Image Processing, VLSI design and so on. He has many publications which are published many good quality journals.

**M. M. A. Hashem** received the Bachelors degree in Electrical and Electronic Engineering from Khulna University of Engineering and Technology (KUET), Khulna, Bangladesh in 1988, Masters degree in Computer Science from Asian Institute of Technology (AIT), Bangkok, Thailand in 1993 and PhD degree in Artificial Intelligence Systems from the Saga University, Japan in 1999. He is currently a Professor in the Dept. of Computer Science and Engineering, Khulna University of Engineering and Technology (KUET), Bangladesh. His research interest includes Evolutionary Computations, Intelligent Computer Networking, Wireless Networking, Soft Computing, Evolutionary Cluster Applications to Evolutionary Robots, Series: Studies in Fuzziness and Soft Computing, Vol. 147, Springer Verlag, Berlin/New York, ISBN: 3540-20901-8, (2004).